\RequirePackage{ifpdf}
\ifpdf 
\documentclass[pdftex]{sigma}
\else
\documentclass{sigma}
\fi

\numberwithin{equation}{section}

\renewcommand\t{\tilde}

\begin{document}

\allowdisplaybreaks

\renewcommand{\PaperNumber}{093}

\FirstPageHeading

\ShortArticleName{From $sl_q(2)$ to a Parabosonic Hopf Algebra}

\ArticleName{From $\boldsymbol{sl_q(2)}$ to a Parabosonic Hopf Algebra}

\Author{Satoshi TSUJIMOTO~$^\dag$, Luc VINET~$^\ddag$ and Alexei ZHEDANOV~$^\S$}

\AuthorNameForHeading{S.~Tsujimoto, L.~Vinet and A.~Zhedanov}

\Address{$^\dag$~Department of Applied Mathematics and Physics, Graduate
School of Informatics,\\
\hphantom{$^\dag$}~Kyoto University, Sakyo-ku, Kyoto 606--8501, Japan}
\EmailD{\href{mailto:tujimoto@amp.i.kyoto-u.ac.jp}{tujimoto@amp.i.kyoto-u.ac.jp}}

\Address{$^\ddag$~Centre de Recherches Math\'ematiques, Universit\'e de Montr\'eal,\\
\hphantom{$^\ddag$}~P.O.~Box 6128, Centre-ville Station, Montr\'eal (Qu\'ebec), H3C 3J7 Canada}
\EmailD{\href{mailto:luc.vinet@umontreal.ca}{luc.vinet@umontreal.ca}}

\Address{$^\S$~Donetsk Institute for Physics and Technology, Donetsk 83114,  Ukraine}
\EmailD{\href{mailto:zhedanov@fti.dn.ua}{zhedanov@fti.dn.ua}}

\ArticleDates{Received August 25, 2011;  Published online October 07, 2011}

\Abstract{A Hopf algebra with four generators among which an involution
(ref\/lection) operator, is introduced. The def\/ining relations
involve commutators and anticommutators. The discrete series
representations are developed. Designated by $sl_{-1}(2)$, this
algebra encompasses the Lie superalgebra $osp(1|2)$. It is
obtained as a $q=-1$ limit of the $sl_q(2)$ algebra  and seen to
be equivalent to the parabosonic oscillator algebra in irreducible
representations. It possesses a noncocommutative coproduct. The
Clebsch--Gordan coef\/f\/icients (CGC) of $sl_{-1}(2)$ are obtained and
expressed in terms of the dual $-1$ Hahn polynomials. A~generating
function for the  CGC is derived using a Bargmann realization.}


\Classification{17B37; 17B80; 33C45}

\section{Introduction}
On the one hand, algebraic structures are
natural descriptors of symmetries. On the other, the exact
solutions of the dynamical equations of physical systems, when
they exist, are typically presented in terms of special functions
and orthogonal polynomials. Not surprisingly hence, the relations
between solvable models, special functions, symmetries and their
algebraic translations is of considerable interest.

The presence of ref\/lection operators has been seen to arise in
many contexts, physical and mathematical, related in particular,
to the f\/irst two of the above areas. To give some examples, recall
that in integrable many-body problems of the Calogero type,
operators with ref\/lections play a key role in expressing the
constants of motion that are in involution \cite{LL,Dunkl_C}. There is currently much activity also in the study
of Dunkl harmonic analysis~\cite{Ros}.

Recently, we have examined univariate polynomials that are
eigenfunctions of operators of Dunkl type, that is of operators
that are f\/irst order in the derivative and involve ref\/lections. We
have thus discovered certain families of ``classical'' orthogonal
polynomials that had hitherto escaped notice \cite{VZ_big,VZ_Bochner}.

It has been found that these polynomials can be identif\/ied as a $q
\to -1$ limits of some $q$-orthogonal polynomials, the simplest
among them being the little $-1$ Jacobi polynomials introduced in
\cite{VZ_little}.

In \cite{BI} and \cite{-1_Hahn} this approach was generalized to
Dunkl shift operators. This provided a~theo\-retical framework for
the Bannai--Ito and the dual $-1$ Hahn polynomials.

With this perspective, it is thus natural to examine algebraic
structures involving ref\/lection operators and it is the purpose of
this paper to contribute to such a study. For related
investigations see, e.g.~\cite{DKT, KD, JSJ,JSJ2}.

\section[Definition of the $sl_{-1}(2)$ algebra and its relation with the $osp(1|2)$ Lie superalgebra]{Def\/inition of the $\boldsymbol{sl_{-1}(2)}$ algebra and its relation\\ with the $\boldsymbol{osp(1|2)}$ Lie superalgebra}

We def\/ine $sl_{-1}(2)$ as the algebra
which is generated by the four elements $J_0$, $J_{\pm}$ and $R$
subject to the relations
\begin{gather}
 [J_0,J_{\pm}]= \pm J_{\pm}, \qquad
[R, J_0]=0, \qquad \{J_{+}, J_{-} \}=2 J_0, \qquad \{R , J_{\pm}\}
=0, \label{comm_J}
\end{gather} where $[A,B]=AB-BA$ and $\{A,B\}=AB+BA$. The
operator $R$ is an involution operator, i.e.\ it satisf\/ies the
property
\begin{gather*}
 R^2=I . 
 \end{gather*}

The Casimir operator $Q$, which by def\/inition commutes with all
the generators ($R, J_0, J_{\pm}$), is
\begin{gather} Q=J_{+} J_{-}R
-\left(J_0 -1/2 \right)R . \label{Q}
\end{gather}

Like the ordinary $sl(2)$ or its quantum analogue $sl_q(2)$, the
algebra $sl_{-1}(2)$ possesses a nontrivial discrete series
representation.

Indeed, let $e_n$, $n=0,1,2,\dots$ denote the basis vectors, and
def\/ine the action of the operators by the formulas
\begin{gather*} J_0 e_n =
(n+ \mu+1/2) e_n, \qquad J_{-} e_n = \rho_n e_{n-1} , \qquad J_{+}
e_n = \rho_{n+1} e_{n+1}, 
\end{gather*} where $\mu$ is a constant
and $\rho_n$ are the positive matrix elements of the
representation. Moreover, demand that $\rho_0=0$ in order to
obtain the standard discrete series bounded from below and with
$n=0,1,2,\dots$.

The operator $R$ commutes with $J_0$ and hence can be diagonalized
in the basis~$e_n$. A simple analysis based on the properties of
$R$, leads to the conclusion that
\begin{gather} R e_n = \epsilon (-1)^n e_n,
\qquad n=0,1,2,\dots, \label{Re}
\end{gather} where $\epsilon = \pm 1$ is a
f\/ixed parameter in a given representation.

Expressing the commutation relations in the basis $e_n$ gives the
following equation for $\rho_n$
\begin{gather*}
 \rho_n^2 + \rho_{n+1}^2 =
2(n+\mu+1/2) 
\end{gather*} with general solution
\begin{gather*} \rho_n^2 =
n+\mu + \kappa(-1)^n, 
\end{gather*} where $\kappa$ is an
arbitrary constant.

The condition $\rho_0=0$ means that $\kappa= -\mu$ and we thus
have \begin{gather*} \rho_n^2 =  n+\mu(1-(-1)^n). 
\end{gather*} The
Casimir operator \eqref{Q}, as should be, is a multiple of the
identity operator{\samepage \begin{gather*} Q e_n = -\epsilon \mu e_n 
\end{gather*} on
the module with the basis $\{e_n\}$.}

The matrix elements can be presented in the form
\begin{gather*} \rho_n^2 = n+
\mu(1-(-1)^n)=[n]_{\mu}, 
\end{gather*} in terms of the
``mu-numbers''
\begin{gather} [n]_{\mu} = n+ \mu(1-(-1)^n). \label{mu_num}
\end{gather} We
def\/ine also the ``mu-factorials'' by
\begin{gather*} [n]_{\mu}! =
[1]_{\mu}[2]_{\mu}[3]_{\mu} \cdots [n]_{\mu}. 
\end{gather*}

If we assume that
\begin{gather*} \mu>-1/2 
\end{gather*} then $\rho_n^2>0$
for $n=1,2,3,\dots$, and we thus obtain a unitary
inf\/inite-dimensional representation of the algebra $sl_{-1}(2)$.
The value of the Casimir operator is $Q=-\epsilon \mu$ in this
representation.

Thus, the discrete series representation is f\/ixed by two
parameters $\epsilon = \pm 1$ and $\mu>-1/2$.

Let us now indicate the connection that $sl_{-1}(2)$ has with the
simplest Lie superalgebra $osp(1|2)$. Consider the elements
$K_{\pm}= J_{\pm}^2$. It is easy to verify that $J_0$, $K_{+}$ and
$K_-$ satisfy together the commutation relations of the $sl(2)$
algebra \begin{gather*} [K_{-},K_{+}]=4 J_0, \qquad [J_0,K_{\pm}]=\pm 2
K_{\pm}. 
\end{gather*} Hence, $J_0$, $J_{\pm}$, $K_{\pm}$ form a
basis for the Lie superalgebra $osp(1|2)$~\cite{FSS}. The
operators $J_0$, $K_{\pm}$ belong to the even part of this algebra,
while the operators $J_{\pm}$ belong to the odd part.

The Casimir operator \eqref{Q} of the $sl_{-1}(2)$ algebra contains
the involution operator $R$ which commutes with the operators
$J_0$ and $J_{+}J_{-}$. Hence the square $Q^2$ of the Casimir
operator will commute with all the generators of the $sl_{-1}(2)$
algebra. However its expression will contain only the operators
$J_0, J_{\pm}$ and not $R$:
\begin{gather*} Q^2 = (J_0-1/2)^2 - J_{+}^2J_{-}^2
- J_{+}J_{-} = (J_0-1/2)^2 - K_{+}K_{-} - J_{+}J_{-}. 
\end{gather*}
This operator coincides with the Casimir operator of the Lie
superalgebra $osp(1|2)$~\cite{FSS}. We see that the Casimir
operator $Q$ of the algebra $sl_{-1}(2)$ can be considered as a
``square root'' of the Casimir operator for the algebra $osp(1|2)$.

In the next section we show that the algebra $sl_{-1}(2)$ can be
obtained as a $q \to -1$ limit of the algebra $sl_q(2)$. This
justif\/ies the name of the algebra.

\section[The $sl_{-1}(2)$ algebra as a limit of the $sl_q(2)$ algebra]{The $\boldsymbol{sl_{-1}(2)}$ algebra as a limit of the $\boldsymbol{sl_q(2)}$ algebra}

 Consider the algebra generated by three
operators $J_0$, $J_{\pm}$, with commutation relations \cite{FV}
\begin{gather}
[J_0, J_{\pm}]=\pm J_{\pm}, \qquad J_{-} J_{+} - qJ_{+}J_{-} = 2
\frac{q^{2J_0}-1}{q^2-1}, \label{J_slq}
\end{gather} where $q$ is a real
parameter.

The Casimir operator $Q$, commuting with $J_0$ and $J_{\pm}$ is
\begin{gather*} Q=J_{+}J_{-} q^{-J_0} -\frac{2}{(q^2-1)(q-1)}
\big(q^{J_0-1}+q^{-J_0}\big). 
\end{gather*} In what follows we restrict
ourselves to discrete series representations of the algebra~\eqref{J_slq}. This means representations that have bases $e_n$, $
n=0,1,\dots$ such that
\begin{gather*} J_0 e_n = (n+\nu) e_n, \qquad J_{-} e_n
= r_n e_{n-1}, \qquad J_{+} e_{n+1} = r_{n+1} e_{n+1}. 
\end{gather*}
As usual, the condition $r_0=0$ is assumed. It is easily verif\/ied
that
\begin{gather*} r^2_n= \frac{2(1-q^n)(1-q^{n+2\nu-1})}{(q+1)(q-1)^2}.
\end{gather*} The parameter $\nu$ is related to the value of
the Casimir operator
\begin{gather*} Q=\frac{2(q^{\nu-1} +
q^{-\nu})}{(1-q)(q^2-1)} 
\end{gather*} in these representations.
The Fock--Bargmann realization of the algebra \eqref{J_slq} can be
def\/ined on the space of polynomials in the variable~$z$ by the
formulas:
\begin{gather} J_0 = z
\partial_z + \nu, \qquad J_{+} = z, \qquad J_{-} = \alpha z
D_q^2 + \beta D_q, \label{FB_slq}
\end{gather} where
\begin{gather*}
\alpha=\frac{2q^{2\nu}}{1+q}, \qquad \beta =
\frac{2(1-q^{2\nu})}{1-q^2} 
\end{gather*} and $D_q$ is the
standard $q$-derivative operator
\begin{gather*} D_q f(z) = \frac{f(zq) -
f(z)}{z(q-1)}. 
\end{gather*} In this realization the basis vectors
$e_n(z)$ are the monomials $e_n(z) = \gamma_n z^n$, where
\begin{gather*}
\gamma_n= \frac{1}{\sqrt{r_1 r_2 \cdots r_n}} 
\end{gather*} is the
normalization coef\/f\/icient.

When $0<q<1$, the algebra def\/ined by \eqref{J_slq} is equivalent to
the quantum $sl_q(2)$ algebra def\/ined by the relations
\begin{gather*}
[A_0,
A_{\pm}]=\pm A_{\pm}, \qquad [A_{-} ,A_{+}]  = 2  \frac{q^{A_0}
-q^{-A_0}}{q-q^{-1}}. 
\end{gather*} Indeed, under the
identif\/ications
\begin{gather}
J_{+}=A_{+}q^{(A_0-1)/2}, \qquad
J_{-}=q^{(A_0-1)/2}A_{-}, \qquad J_0=A_0, \label{JA_cor}
\end{gather} the
commutation relations~\eqref{J_slq} are transformed into the
commutation relations~\eqref{JA_cor}.

When $q \to 1$ the algebra $sl_q(2)$ with the def\/ining relations
\eqref{J_slq} becomes the $sl(2)$ algebra:
\begin{gather*} [J_0, J_{\pm}]=\pm
J_{\pm}, \qquad [J_{-}, J_{+}] = 2J_0. 
\end{gather*}

There is also a nontrivial limit when $q \to -1$. It is obvious
that the commutation relations~\eqref{J_slq} become the commutation
relations~\eqref{comm_J} when $q=-1$. The limit process for the
matrix coef\/f\/icients~$r_n$ is more subtle however.

Assume that $\nu=j=1,2,3,\dots$ is a positive integer. Let
$q=-e^{\tau}$, then the limit $q \to -1$ is equivalent to the
limit $\tau \to 0$.

Assume f\/irst that $n=0,2,4,\dots$ is even. Then
\[
r_n^2=\frac{2(1-(-1)^n e^{\tau n})(1-(-1)^{n+2\nu-1}
e^{\tau(n+2\nu-1)})}{(1-e^{\tau})(1+e^{\tau})^2} =\frac{2(1-
e^{\tau n})(1+e^{\tau(n+2\nu-1)})}{(1-e^\tau)(1+e^\tau)^2}.
\]
Hence
\[
\lim_{q \to -1} r_n^2 =  \lim_{\tau \to 0} \frac{1- e^{\tau
n}}{1-e^\tau} =n.
\]
When $n$ is odd, we have
\[
r_n^2 =\frac{2(1+ e^{\tau n})(1-
e^{\tau(n+2\nu-1)})}{(1-e^\tau)(1+e^\tau)^2},
\]
hence
\[
\lim_{q \to -1} r_n^2 =  \lim_{\tau \to 0} \frac{1- e^{\tau
(n+2\nu -1)}}{1-e^\tau} =n+2\nu-1,
\]
and
\[
\lim_{q \to -1} r_n^2 = n + \mu(1-(-1)^n) =[n]_{\mu} = \rho_n^2
\]
where $\nu = \mu+1/2$.

Thus, for integer values of the parameter $\nu$ the limit $q \to
-1$ of the matrix elements $r_n$ gives the expected matrix
elements $\rho_n$ of the discrete series of the $sl_{-1}(2)$
algebra.

When $\nu$ is not an integer, the limit of $r_n$ is not well
def\/ined. In this case we can assume that the limiting matrix
element $\rho_n^2$ is obtained by a linear interpolation from the
integer $\nu$ case.

If $\nu=j$ is integer, the involution operator $R$ can also be
obtained in the limit $q \to -1$
\[
R = \lim_{q \to -1} q^J_0.
\]
Indeed, we have
\[
q^{J_0} e_n = q^{n+j} e_n.
\]
So, in the limit
\[
R e_n = \epsilon (-1)^n e_n,
\]
where
\[
\epsilon= \lim_{q \to -1} q^{j} = (-1)^{j} = \pm 1.
\]
This uniquely characterizes the involution operator with the
property~$R^2=I$.

We thus see that the generators $J_0, J_{\pm 1}$ and $R$ of
$sl_{-1}(2)$ can be obtained from the al\-gebra~\eqref{J_slq} when the
representation parameter is a positive integer
$\nu=\mu=1,2,3,\dots$. If $\nu$ is a~real positive parameter, then
the limiting process is not well def\/ined and we postulate that in
the limit $q \to -1$ the matrix elements $\rho_n$ correspond to
the matrix elements $r_n$ with~$\nu$ real and positive.

Note that the $q \to -1$ limit considered here is dif\/ferent from
the well known special case of $sl_q(2)$ for q a root of unity~\cite{RA}. In the latter case the operators~$J_{\pm}$ are
nilpotent $J_{\pm}^N=0$, where~$N$ is the order of the root of
unity and hence all irreducible representations are restricted to
be of dimension~$N$. In our case we have inf\/inite-dimensional
representations.

\section[Relation with the parabosonic oscillator and the Fock-Bargmann realization]{Relation with the parabosonic oscillator\\ and the Fock--Bargmann realization}

Consider the commutator $[J_-,J_+]$. We have
\[
[J_-,J_+] = \{J_-,J_+\} -2 J_+J- = 2J_0 - 2 J_+J_- .
\]
Remembering the expression \eqref{Q} for the Casimir operator, we
f\/ind that
\begin{gather*}
[J_-,J_+] = 1-2QR. 
\end{gather*} For
representations with a f\/ixed value $\epsilon = \pm 1$, we have
$Q=-\epsilon \mu$ and hence
\begin{gather} [J_-,J_+] = 1+2\epsilon \mu R .
\label{com_JJ_mu}
\end{gather} This relation, \eqref{com_JJ_mu}, def\/ines the
parabosonic oscillator algebra \cite{Vas, Muk, Mac,Rosen} with opera\-tors~$J_-$, $J_+$, $R$ satisfying the
commutation relations~\eqref{com_JJ_mu} and $\{R, J_{\pm}\}=0$
together with the condition~$R^2=I$.

Conversely, assume that the operators $J_{-}$, $J_{+}$, $R$ form a
representation  of the parabosonic oscillator algebra. We can
def\/ine the operator $J_0$ as $J_0=\{J_{+}, J_{-}\}/2$. Then it is
easily verif\/ied that the operators $J_0$, $J_{+}$, $J_{-}$, $R$ satisfy
the relations~\eqref{comm_J} def\/ining the  $sl_{-1}(2)$ algebra.

Thus, if one restricts to irreducible representations with a f\/ixed
value of the Casimir operator $Q=-\epsilon \mu$, the algebra
$sl_{-1}(2)$ is equivalent to the parabosonic oscillator algebra.

For def\/initeness, in what follows we will use representations for
which  $\epsilon =1$.

We can construct the Fock--Bargmann representation of the
$sl_{-1}(2)$ algebra in terms of f\/irst order
dif\/ferential-dif\/ference operators. Indeed, one can use the well
known realization of the parabosonic operators \cite{Muk,Rosen}
\begin{gather} R = R_x, \qquad J_{+} = x, \qquad J_{-} =
\partial_x + \frac{\mu}{x}(1-R_x), \qquad J_0 = x
\partial_x + \mu+1/2, \label{Fock_J}
\end{gather} where $R_x$ is the ref\/lection
(parity) operator def\/ined by $Rf(x) = f(-x)$ for every function~$f(x)$.

The operator $J_{-}$ coincides in this realization  with the
standard Dunkl operator \cite{Dunkl_C}.

Note that when $\nu$ is integer, the realization \eqref{Fock_J} can
be obtained as a limit $q \to -1$ from the realization~\eqref{FB_slq}.

The basis $e_n(x)$ is here realized by the  monomials
\begin{gather*} e_n(x) =
\gamma_n x^n, 
\end{gather*} with some constants
$\gamma_n$. If we take
\begin{gather*} \gamma_n = \frac{1}{\sqrt{[n]_{\mu}!}},
\end{gather*} we reproduce the canonical formulas of the
previous section
\[
J_{-} e_n(x) = \rho_n e_{n-1} , \qquad J_{+} e_{n}(x) = \rho_{n+1}
e_{n+1}(x).
\]
Sometimes it will be convenient to take $\gamma_n=1$, i.e.\ $e_n(x)
= x^n$. In this case we have
\begin{gather*} J_0 e_n(x) = (n+ \mu+1/2)e_n(x),
\qquad J_{-} e_n(x) = [n]_{\mu} e_{n-1}(x) , \qquad J_{+}e_n(x) =
e_{n+1}(x). 
\end{gather*} Of course, the Casimir operator
reduces (up to a constant factor) to the identity operator
\[
Q e_n(x) = - \mu e_n(x).
\]
Note that similar relations were investigated in \cite{HPV,P1}. Our approach is dif\/ferent, because we start from the
algebra $sl_{-1}(2)$ with 4 generators which is observed to be a
limiting case of the~$sl_q(2)$ algebra. The relation (in
irreducible representations) with the parabose algebra is obtained
a posteriori.

\section[Coproduct and the Clebsch-Gordan coefficients]{Coproduct and the Clebsch--Gordan coef\/f\/icients}

The most important property of the
$sl_{-1}(2)$ algebra is that it admits an ``addition rule'', or a~coproduct which can be inferred from the well known coproduct of
the quantum algebra~$sl_q(2)$.

Assume that we have two independent  representations of the
algebra~\eqref{J_slq} on the linear spaces~$S_1$ and~$S_2$. Let $S_1
\otimes S_2$ be the direct product of these spaces. We will denote
by $A\otimes B$, the direct products of operators acting on the
spaces $S_1$ and $S_2$, $A \in {\rm End}(S_1)$, $B \in {\rm End}(S_2)$. It is
readily verif\/ied that the elements
\begin{gather*} \t J_0=J_0\otimes {I} + {I}
\otimes J_0, \qquad \t J_{\pm } = J_{\pm}\otimes q^{J_0} + {I}
\otimes J_{\pm} 
\end{gather*} again satisfy the commutation
relations~\eqref{J_slq} of the $sl_q(2)$ algebra~\cite{FV}. (Here $I$
stands for the identity operator).

Assuming that the representation parameter $\nu$ is a positive
integer, we have a well-def\/ined  $q \to -1$ limit from $sl_q(2)$
to $sl_{-1}(2)$. The operator $q^{J_0}$ in this limit becomes
$\epsilon R$ with $\epsilon = \pm 1$. It is thus natural to expect
that for arbitrary representation parameter $\mu > -1/2$, the
$sl_{-1}(2)$ algebra admits a coproduct rule.

It can be def\/ined as follows.  For two independent representations
of the $sl_{-1}(2)$ algebra with the Casimir parameters $\mu_1$,
$\mu_2$, let us introduce the following operators $\t J_0$, $\t
J_{\pm}$, $\t R$ that act on the direct product of the spaces~$S_1$,~$S_2$:
\begin{gather} \t J_0=J_0\otimes {I} + {I} \otimes J_0, \qquad \t J_{\pm
} = J_{\pm}\otimes {R} + {I} \otimes J_{\pm}, \qquad \t R = {R}
\otimes {R}. \label{add_rule}
\end{gather} Then the operators $\t J_0$, $\t
J_{\pm}$, $\t R$ satisfy the commutation relations~\eqref{comm_J}, i.e.\
they are again gene\-ra\-tors of the algebra $sl_{-1}(2)$. The
verif\/ication of this statement is elementary.

Note that a similar coproduct  was proposed in~\cite{DKT,KD} for the parabosonic oscillator algebra, in the
identif\/ication of its Hopf algebra structure.

In what follows we restrict ourselves to representations with
$\epsilon_1=\epsilon_2=1$ and $\mu_1>-1/2$, $\mu_2>-1/2$.

In the Fock--Bargmann realization, $S_1$ and $S_2$ are spaces of
polynomials in the arguments, say, $x$ and $y$. We def\/ine
representations with the parameters $\mu_1$ and $\mu_2$ on these
spaces by the formulas
\[
J_0^{(x)} = x \partial_x +\mu_1 +1/2, \qquad J_{+}^{(x)} = x, \qquad
J_{-}^{(x)} = \partial_x + \frac{\mu_1}{x}(1-R_x)
\]
and
\[
J_0^{(y)} = y \partial_y +\mu_2 +1/2, \qquad J_{+}^{(y)} = y, \qquad
J_{-}^{(x)} = \partial_y + \frac{\mu_2}{y}(1-R_y).
\]
The Casimir operators take the constant values $Q_1=-\mu_1$, $Q_2
=-\mu_2$ on these representations.

Following~\eqref{add_rule}, the operators of the coproduct are given
as \begin{gather*} \t J_0 = x
\partial_x + y \partial_y + \mu_1 +\mu_2+1, \qquad \t J_{+} = xR_y+y,\nonumber\\
  \t J_{-} =\big(\partial_x +{\mu_1}{x^{-1}}(1-R_x)\big)R_y + y
\partial_y +{\mu_2}{y^{-1}}(1-R_y). 
\end{gather*} The corresponding
Casimir operator \begin{gather*}
\t Q = \t J_{+} \t J_{-}\t R -\big(\t J_0
-1/2 \big) \t R 
\end{gather*} commutes with the ``local''
Casimir operators $Q_1$ and $Q_2$ and with the operators $\t J_0$,
$\t J_{\pm}$ but not with the operators $J_0^{(x)}$, $J_0^{(y)}$.

Hence we can posit the Clebsch--Gordan problem as follows.

In view of \eqref{add_rule}, the operator $\t J_0$ can take the
eigenvalues $\mu_1+\mu_2+N+1$, where $N=0,1,2,\dots$. We denote by
$\Phi_{N,q}$, the eigenstate with f\/ixed eigenvalues of the total
Casimir operator and of~$\t J_0$:
\begin{gather*} \t Q \Phi_{N,k} =
q_k \Phi_{N,k}, \qquad \t J_{0} \Phi_{N,k} = (\mu_1+\mu_2+N+1)
\Phi_{N,k}. 
\end{gather*} This state can be decomposed as a
linear combination of direct product of states:
\begin{gather} \Phi_{N,k}=
\sum_{s=0}^N W_{s;N,k}  e_s \otimes e_{N-s}, \label{CGd}
\end{gather} with
coef\/f\/icients $W_{s;N,k}$ that can be called the Clebsch--Gordan
coef\/f\/icients of the $sl_{-1}(2)$ al\-gebra.

It is not dif\/f\/icult to see that the Casimir eigenvalue $q_k$ has
the expression
\begin{gather} q_k = (-1)^{k+1} (\mu_1 + \mu_2 +1/2+k), \qquad
k=0,1, \dots, N . \label{q_k}
\end{gather} Indeed, the eigenvalues of the
operator $J_0$ are $n+\mu+1/2=n-\epsilon Q+1/2$ (recall that
$Q=-\epsilon \mu$ in the given representation). Hence, if the
eigenvalue $\lambda>0$ of $J_0$ is f\/ixed, then the possible
eigenvalues of the Casimir operator in absolute value are:
\begin{gather} |Q|
= |\lambda-1/2|,  |\lambda-3/2|,   \dots. \label{abs_Q}
\end{gather}

When considering the coproduct of two $sl_{-1}(2)$ algebras, we
know that the eigenvalues of $\t J_0$ have the form $\t \lambda =
\mu_1+\mu_2+N+1$. Hence, from \eqref{abs_Q} we have for the set of
absolute values (recall that the total number of eigenvalues
should be equal to $N+1$)
\begin{gather*} |\t Q| =\mu_1+\mu_2+N+1/2,
\mu_1+\mu_2+N-1/2, \dots, \mu_1+\mu_2+1/2. 
\end{gather*} The
state with the maximal absolute value $|q_N|=\mu_1+\mu_2+N+1/2$ of
the Casimir opera\-tor~$\t Q$, corresponds to the state $\t e_0$
satisfying the conditions:
\[
\t J_0 \t e_0 =(\mu_1+\mu_2+N+1) \t e_0, \qquad \t J_- \t e_0 =0 .
\]
In order to determine the sign of the eigenvalue $q_N$, we notice
that
\begin{gather} \t R \Phi_{N,k} = (R_1 \otimes R_2) \Phi_{N,k} =(-1)^N
\Phi_{N,k}. \label{R Phi}
\end{gather} This means on the one hand that
\begin{gather*} \t
R \t e_0 = (-1)^N \t e_0. 
\end{gather*}
On the other hand, by
\eqref{Re} $\t R \t e_0 = \t \epsilon  \t e_0$ and hence $\t \epsilon
= (-1)^N$, where $\t \epsilon$ stands for the eigenvalue of the
parity operator $\t R$ on the state $\t e_0$. We thus have
\begin{gather*} q_N
= (-1)^{N+1} (\mu_1+\mu_2+N+1/2). 
\end{gather*} Taking into account
the parity of the coproduct states we arrive at formula~\eqref{q_k}.

In order to f\/ind the coef\/f\/icients $W_{s;N,k}$ we shall derive a
3-term recurrence relation for them.

Taking into account relation \eqref{R Phi}, we see that the
eigenvalue equation $\t Q \Phi_{N,k} = q_k \Phi_{N,k}$ can be
presented in the form
\begin{gather*} Q_0 \Phi_{N,k} = (-1)^N q_k \Phi_{N,k},
\end{gather*} where
\begin{gather*}
 Q_0= \t J_{+}\otimes \t J_{-} - \t J_0 + 1/2. 
 \end{gather*}

From the expression for the Casimir operator it is seen that $Q_0$
is tri-diagonal in the basis $e_s \otimes e_{N-s}$. Hence, the
Clebsch--Gordan coef\/f\/icients $W_{s;N,k}$ satisfy the 3-term
recurrence relation
\begin{gather*} A_{s+1} W_{s+1;N,k} + A_{s} W_{s-1;N,k} +
B_{s} W_{s;N,k} =  (-1)^N q_k W_{s;N,k}, 
\end{gather*} where
the recurrence coef\/f\/icients $A_{s}$, $B_{s}$ are easily expressed in
terms of the known representation matrix elements for
$sl_{-1}(2)$:
\begin{gather*} A_{s}= (-1)^{s} \sqrt{[s]_{\mu_1}
[N-s+1]_{\mu_2} } 
\end{gather*} and
\begin{gather*} B_{s} = (-1)^N \left(
[s]_{\mu_1} + [N-s]_{\mu_2} -N-\mu_1-\mu_2-1/2 \right), 
\end{gather*} where we adopt the notation \eqref{mu_num}.

Note that the expression for the coef\/f\/icient $B_s$ can be
simplif\/ied to:
\begin{gather*} B_s= \begin{cases} -\frac{1}{2}-(-1)^s(\mu_1+\mu_2)
\quad \mbox{if} \quad N \quad \mbox{even},\\
\frac{1}{2}+(-1)^s(\mu_1-\mu_2) \quad \mbox{if} \quad N \quad
\mbox{odd}.\end{cases}
\end{gather*}

Thus the CGC are expressed in terms of some orthogonal polynomials
$P_s(x)$
\begin{gather}
W_{s;N,k}= W_{0;N,k} P_s(q_k;N). \label{W_P}
\end{gather} These
orthogonal polynomials satisfy the 3-term recurrence relation
\begin{gather*}
A_{s+1} P_{s+1}(x) + A_s P_{s-1}(x) + B_s P_s(x) = x P_{s}(x)
\end{gather*} with initial conditions $P_{-1}=0$, $P_0=1$.
From the above expressions for $A_s$, $B_s$ we can conclude that the
polynomials $P_s(x)$ coincide with the generic dual $-1$ Hahn
polynomials~\cite{-1_Hahn}.

Indeed, it is convenient to present the polynomials $P_n(x)$ in
monic form
\[
P_n(x) = \frac{\hat P_n(x)}{A_1 A_2 \cdots A_n}.
\]
Then the polynomials $\hat P_n(x) = x^n + O(x^{n-1})$ satisfy on
the one hand
\begin{gather*}
\hat P_{n+1}(x) + u_n \hat P(x) + B_n \hat P_n(x)
= x\hat P_n(x), 
\end{gather*} where
\begin{gather*} u_n = A_n^2
=[n]_{\mu_1} [N-n+1]_{\mu_2}. 
\end{gather*} Note that $u_n >0$,
$n=1,2,\dots, N$ and $u_{N+1}=0$.

On the other hand, the dual $-1$ Hahn polynomials \cite{-1_Hahn}
$R_n^{(-1)}(x;\alpha,\beta;N)$ depend on 3 para\-me\-ters
$\alpha$, $\beta$ and $N=1,2,\dots$ and obey the recurrence relation
\begin{gather*}
R_{n+1}^{(-1)}(x) + u_n^{(-1)} R_{n-1}^{(-1)}(x) + b_n^{(-1)}
R_n^{(-1)}(x) = x R_n^{(-1)}(x), 
\end{gather*} where the
recurrence coef\/f\/icients are \cite{-1_Hahn}
\begin{gather*} u_n=
4[n]_{\xi}[N+1-n]_{\eta}, \qquad b_n = 2([n]_{\xi} + [N-n]_{\eta})
+\zeta .
\end{gather*}
The parameters $\xi$, $\eta$, $\zeta$ are
related to the parameters $\alpha$, $\beta$, $N$. When $N$ is even
\begin{gather}
\xi=\frac{\beta-N-1}{2}, \qquad \eta=\frac{\alpha-N-1}{2}, \qquad
\zeta=1-\alpha-\beta. \label{N_even_xi}
\end{gather}
When $N$ is odd  \begin{gather}
\xi=\alpha/2, \qquad \eta=\beta/2, \qquad \zeta=-2N-1-\alpha-\beta.
\label{N_odd_xi}
\end{gather}
Comparing the recurrence coef\/f\/icients of the
polynomials $\hat P_n(x)$ with the corresponding coef\/f\/i\-cients of
the dual $-1$ Hahn polynomials we conclude that
\begin{gather*} \hat P_n(x)=
2^{-n} R_n^{(-1)}(2(x-x_0); \alpha, \beta,N), 
\end{gather*} where
the parameters $\alpha$, $\beta$ are found from formulas
\eqref{N_even_xi} and \eqref{N_odd_xi} with $\xi=\mu_1$, $\eta=\mu_2$.
The shift parameter $x_0$ can also be expressed in terms of
$\mu_1$, $\mu_2$ in an obvious way.

We thus expressed the Clebsch--Gordan coef\/f\/icients of the
$sl_{-1}(2)$ algebra in terms of the dual $-1$ Hahn polynomials
$R_n^{(-1)}(x;\alpha,\beta;N)$.

The remaining problem is to f\/ind an explicit expression for the
coef\/f\/icient $W_{0;N,k}$ in~\eqref{W_P}. This can be done using the
following observation. The vectors $\psi_s = e_s \otimes e_{N-s}$
form an orthonormal  basis in the $N+1$-dimensional linear space.
There is thus a scalar product such that
\begin{gather*} (\psi_s, \psi_t) =
\delta_{st}. 
\end{gather*} The vectors $\Phi_{N,k}$ form
another orthonormal basis on the same space and so:
\begin{gather*}
(\Phi_{N;k}, \Phi_{N;l})=\delta_{kl}. 
\end{gather*}
Hence, the
matrix $W_{s;N,k}$ is orthogonal, i.e.\ it obeys
\begin{gather*} \sum_{k=0}^N
W_{n;N,k} W_{m;N,k} = \delta_{nm}. 
\end{gather*} Taking into
account formula \eqref{W_P} we thus have on the one hand
\begin{gather*}
\sum_{k=0}^N W_{0;N,k}^2 P_n(q_k) P_m(q_k) = \delta_{nm}.
\end{gather*}
On the other hand, the orthonormal dual $-1$ Hahn
polynomials $P_n(x)$ satisfy the orthogonality property
\cite{-1_Hahn}
\begin{gather*} \sum_{k=0}^N w_k P_n(q_k) P_m(q_k) =
\delta_{nm}, 
\end{gather*} where $w_k$ are positive discrete
weights (concentrated masses) localized at the spectral points
$q_k$. (The positivity property $w_k>0$ follows from the
positivity of the recurrence coef\/f\/icients $u_n>0$, $n=1,2,\dots,N$~\cite{-1_Hahn}.)

We thus have \begin{gather*}
W_{0;N,k} = \sqrt{w_k}. 
\end{gather*} Explicit
expressions for the weights were found in~\cite{-1_Hahn}. This
solves the problem of f\/inding the Clebsch--Gordan coef\/f\/icients
$W_{s;N,k}$ up to sign factors~$\pm 1$.

The result is not surprising. We have seen that the $sl_{-1}(2)$
algebra is a $q \to -1$ limit of the $sl_q(2)$ algebra and for the
latter algebra, the CGC are expressed in terms of the dual $q$-Hahn
polynomials~\cite{KK}.

Also, when $\mu_1=\mu_2=0$ the dual $-1$  Hahn polynomials coincide
with the ordinary Kraw\-tchouk polynomials. This result is also
expected: the case  $\mu_1=\mu_2=0$ corresponds to the case when
both $sl_{-1}(2)$ algebras in the product are equivalent to
oscillator algebras whose Clebsch--Gordan coef\/f\/icients are
expressed in terms of Krawtchouk polynomials~\cite{VK}. Note
nevertheless, that even if we start with pure oscillator algebras
(i.e.~$\mu_1=\mu_2=0$), the addition rule is non-standard: it
involves the ref\/lection operator. Hence even in this simplest case
the composed algebra will not be a pure oscillator algebra.

\section[The Clebsch-Gordan problem in the Fock-Bargmann picture]{The Clebsch--Gordan problem in the Fock--Bargmann picture}

The Clebsch--Gordan problem can be
considered also in the Fock--Bargmann picture. This leads to a~generating function for the Clebsch--Gordan coef\/f\/icients.

The representation space  for the coproduct is the space of
polynomials in two variables $f(x,y)$ which are homogeneous of
degree $N$:
\begin{gather} f(x,y) = y^N \Phi(x/y), \label{f_psi_N}
\end{gather} where
$\Phi(z)$ is a polynomial of degree $N$ in the variable $z$.

For f\/ixed $N$ the action of the operator operator $\t J_0$ is
diagonal: it has the eigenvalue $N+\mu_1+\mu_2+1$ (due to Euler's
theorem on homogeneous polynomials).

Using the representation \eqref{f_psi_N}, we obtain the eigenvalue
equation
\begin{gather} \t Q f(x,y) = q_k f(x,y), \label{Qqf}
\end{gather} where the
eigenvalues $q_k$ are given by \eqref{q_k}.

Substituting $f(x,y)$ expressed as in \eqref{f_psi_N} into \eqref{Qqf}
we obtain a dif\/ferential-dif\/ference equation for the function
$\Phi(z)$:
\begin{gather} L \Phi_k(z) = q_k \Phi_k(z), \label{L_Phi}
\end{gather} where
the operator $L$ is
\begin{gather} L= (-1)^N (z^2+1) \partial_z R  +\left(
(-1)^N\frac{\mu_1}{z}  - (-1)^N (\mu_2+N)z - \mu_1 -(-1)^N
\mu_2\right) R \nonumber\\
\phantom{L=}{}+ \left(\mu_2 z - \frac{1}{2} -
(-1)^N \frac{\mu_1}{z} \right) I, \label{L_N}
\end{gather} and where $R$ acts
according to $R\Phi(z) = \Phi(-z)$ and $I$ is the identity
operator.

The operator $L$ preserves the linear space of polynomials of
degree $\le N$ and  belongs to a~class of Dunkl type operators of
the f\/irst order considered in \cite{VZ_little, VZ_big,VZ_Bochner}. More precisely, the opera\-tor~$L$ is a linear
combination (with coef\/f\/icients depending on $z$) of the operators~$I$,~$R$ and~$\partial_z R$.  The main dif\/ference with respect to
the Dunkl type operators used in the papers mentioned above is
that the operator~\eqref{L_N} does not preserve the whole space of
polynomials of a given arbitrary degree. Moreover, it is seen that
the operator~\eqref{L_N} is 3-diagonal in the monomial basis~$z^n$, $n=0,1,\dots, N$.

Using the decomposition of the function $\Phi(z) = \Phi_{\rm e}(z) +
\Phi_{\rm o}(z)$ into its even~$\Phi_{\rm e}(z)$ and odd~$\Phi_{\rm o}(z)$ parts we
can reduce the equation~\eqref{L_Phi} to standard hypergeometric
equations for~$\Phi_{\rm e}(z)$ and~$\Phi_{\rm o}(z)$.

The explicit form of the solution will depend on the parity of the
integers $N$ and $k$.

When both $N$ and $k$ even, we have
\begin{gather*}
\Phi_k(z) = {_2}F_1
\left( {-\frac{k}{2}, -\mu_2 -\frac{k}{2} + \frac{1}{2} \atop
\mu_1 + \frac{1}{2} } ; -z^2 \right) \big(1+z^2\big)^{\frac{N-k}{2}}  \nonumber \\
\phantom{\Phi_k(z) =}{}+ \frac{kz}{2 \mu_1+1} \;   {_2}F_1 \left(
{1-\frac{k}{2}, -\mu_2 -\frac{k}{2} + \frac{1}{2} \atop \mu_1 +
\frac{3}{2} } ; -z^2 \right) \big(1+z^2\big)^{\frac{N-k}{2}}, 
\end{gather*}
when $N$ is even and $k$ is odd
\begin{gather*}
\Phi_k(z) = {_2}F_1 \left(
{-\frac{k+1}{2}, -\mu_2 -\frac{k}{2} \atop \mu_1 + \frac{1}{2} } ;
-z^2 \right) \big(1+z^2\big)^{\frac{N-k-1}{2}} \nonumber \\
\phantom{\Phi_k(z) =}{}- \frac{2 \mu_1+2\mu_2+k+1}{2 \mu_1+1}   z  \; {_2}F_1 \left(
{-\frac{k-1}{2}, -\mu_2 -\frac{k}{2}  \atop \mu_1 + \frac{3}{2} }
; -z^2 \right) \big(1+z^2\big)^{\frac{N-k-1}{2}}, 
\end{gather*} for $N$ odd
and $k$ even we have
\begin{gather*}
\Phi_k(z) = {_2}F_1 \left(
{-\frac{k}{2}, -\mu_2 -\frac{k+1}{2} \atop \mu_1 + \frac{1}{2} } ;
-z^2 \right) \big(1+z^2\big)^{\frac{N-k-1}{2}}  \nonumber
\\
\phantom{\Phi_k(z) =}{}
+\frac{2\mu_1+k+1}{2 \mu_1+1}   z \: {_2}F_1 \left(
{-\frac{k}{2}, -\mu_2 -\frac{k-1}{2}  \atop \mu_1 + \frac{3}{2} }
; -z^2 \right) \big(1+z^2\big)^{\frac{N-k-1}{2}}, 
\end{gather*}  f\/inally for
$N$ and $k$ odd
\begin{gather*}
 \Phi_k(z) = {_2}F_1 \left( {-\frac{k-1}{2},
-\mu_2 -\frac{k}{2} \atop \mu_1 + \frac{1}{2} } ; -z^2 \right)
\big(1+z^2\big)^{\frac{N-k}{2}}   \nonumber \\
\phantom{\Phi_k(z) =}{}- \frac{2 \mu_2+k}{2
\mu_1+1}   z \; {_2}F_1 \left( {-\frac{k-1}{2}, 1-\mu_2
-\frac{k}{2} \atop \mu_1 + \frac{3}{2} } ; -z^2 \right)
\big(1+z^2\big)^{\frac{N-k}{2}}. 
\end{gather*} (All these functions
$\Phi_k(z)$ are def\/ined  up to a common normalization factor.)

The solutions $\Phi_k(z)$ are polynomials of degree $N$ in $z$. It
is clear from the def\/inition \eqref{CGd} and \eqref{f_psi_N} that the
series expansion \begin{gather*}
\Phi_k(z) = \sum_{s=0}^N C_s^{(k)} z^s
\end{gather*} gives the Clebsch--Gordan coef\/f\/icients
\[
C_s^{(k)} = W_{s;N,k}.
\]
The polynomials $\Phi_k(z)$ are thus generating functions for the
Clebsch--Gordan coef\/f\/icients of the $sl_{-1}(2)$ algebra and hence
for the dual $-1$ Hahn polynomials.

\subsection*{Acknowledgments}

The authors are grateful to M.S.~Plyushchay for drawing their
attention to~\cite{HPV} and~\cite{P2}. The authors would
like to gratefully acknowledge the hospitality extended to LV and
AZ by Kyoto University and to ST and LV by the Donetsk Institute
for Physics and Technology  in the course of this investigation.
The research of ST is supported
in part through funds provided by KAKENHI (22540224), JSPS.
The research of LV is supported in part by a research grant from
the Natural Sciences and Engineering Research Council (NSERC) of
Canada.

\pdfbookmark[1]{References}{ref}
\LastPageEnding

\end{document}